\documentclass[12pt]{article}
\usepackage{amsmath}
\usepackage{graphicx,psfrag,epsf}
\usepackage{enumerate}
\usepackage{natbib}
\usepackage{url} 
\def\bSig\mathbf{\Sigma}

\usepackage{natbib}
\usepackage{amsmath}
\usepackage{graphicx}
\usepackage{rotating}
\usepackage{natbib}
\usepackage{url} 
\usepackage[utf8]{inputenc}
\usepackage{amsmath}
\usepackage{graphicx}
\usepackage{enumerate}
\usepackage{pdflscape,multirow}

\usepackage{url} 
\usepackage[utf8]{inputenc} 
\usepackage{graphicx}
\usepackage{xcolor,yfonts,eufrak}
\usepackage{blkarray}
\usepackage{amsmath,amssymb,amsthm,amsfonts,natbib,bbm,bm,color,setspace,graphics,graphicx,indentfirst,url,caption,multicol,multirow,longtable,tabu,verbatim,enumerate,footnote,multirow}
\usepackage{tikz,tikzscale}
\usepackage{pifont}
\usetikzlibrary{fit,positioning}
 \usetikzlibrary{bayesnet, arrows}
\usepackage{graphicx}
 \usepackage{multirow}
 \usepackage{lscape}
\usepackage{tablefootnote}
\usepackage{ulem}
\usepackage{amssymb}
\usepackage{algorithm}
\usepackage{algpseudocode}
\usepackage{xcolor}
\usepackage{soul}
\usepackage{amsmath}
\usepackage{enumerate}
\usepackage{natbib} 

\newcommand{\blind}{1}

\def\spacingset#1{\renewcommand{\baselinestretch}%
{#1}\small\normalsize}

\begin{document}

\spacingset{1}

 
\if1\blind
{
  \title{\bf A Bayesian Zero-Inflated Dirichlet-Multinomial Regression Model for Multivariate Compositional Count Data}
  \author{Matthew D. Koslovsky \thanks{  Department of Statistics, Colorado State University, Fort Collins, CO, USA, email: matt.koslovsky@colostate.edu  }
    }
  \maketitle
} \fi

\if0\blind
{
  \bigskip
  \bigskip
  \bigskip
  \begin{center}
    {\LARGE\bf Title}
\end{center}
  \medskip
} \fi

\bigskip
\begin{abstract}
The Dirichlet-multinomial (DM) distribution plays a fundamental role in modern statistical methodology development and application. Recently, the DM distribution and its variants have been used extensively to model multivariate count data generated by high-throughput sequencing technology in omics research due to its ability to accommodate the compositional structure of the data as well as overdispersion. A major limitation of the DM distribution is that it is unable to handle excess zeros typically found in practice which may bias inference. To fill this gap, we propose a novel Bayesian zero-inflated DM model for multivariate compositional count data with excess zeros. We then extend our approach to regression settings and embed sparsity-inducing priors to perform variable selection for high-dimensional covariate spaces. Throughout, modeling decisions are made to boost scalability without sacrificing interpretability or imposing limiting assumptions. Extensive simulations and an application to a human gut microbiome data set are presented to compare the performance of the proposed method to existing approaches. We provide an accompanying \texttt{R} package with a user-friendly vignette to apply our method to other data sets. 
\end{abstract}

\noindent%
\vfill

\newpage
\spacingset{1.45} 

\section{Introduction}\label{Intro}
 
 \textcolor{black}{
 The human microbiome is the collection of microorganisms that live on and inside of our bodies. A major aim in human microbiome studies is investigating the feasibility of designing personalized dietary interventions that modulate and maintain the composition of the microbiome to diagnose and treat microbiome-associated diseases \citep{xu2015dietary}. Despite recent technological and computational advances for human microbiome research, efficacious intervention strategies require a deeper understanding of the dietary factors associated with the composition and function of a healthy microbiome \citep{johnson2019daily}. The methodological developments proposed in this work were motivated by data collected in the Cross-sectional Study of Diet and Stool Microbiome Composition (COMBO), which was designed to explore dietary patterns linked to gut microbial enterotypes \citep{wu2011linking}. Analyzing these data is challenged by the large number of potential associations between each dietary factor and each microbial taxon, as well as the compositional structure of the data, overdispersion, and zero-inflation, characteristic of microbiome samples. Our objective is to develop a novel Bayesian zero-inflated Dirichlet-multinomial model to estimate microbial relative abundances and explore the relation between exogenous and endogenous factors and microbial composition in the presence of excess zeros without sacrificing interpretability or imposing limiting assumptions that may bias inference. Our approach differs from existing methods as it simultaneously estimates individual- and population-level microbial abundances, quantifies parameter uncertainty, is able to accommodate and identify covariates associated with microbial abundances as well as potential zero-inflation, and is scalable to the large covariate and compositional spaces encountered in practice.}

\subsection*{Related Work}
The Dirichlet-multinomial (DM) distribution plays a fundamental role in modern statistical methodology development and application. Recently, the DM distribution and its variants have been used extensively to model multivariate count data generated by high-throughput sequencing technology in omics research due to its ability to accommodate the compositional structure of the data (i.e., the magnitude of a single component depends on the sum of all the components’ counts) as well as overdispersion. 
A seemingly inconsequential characteristic of the DM distribution is that estimated probabilities for zero counts are strictly positive, even if the true probability of occurrence is zero. While oftentimes overlooked in practice, this limitation has profound implications on modeling and inference (see the Supporting Information for a toy example demonstrating the impacts of ignoring zero-inflation on inference). 


Typically, zero-inflated models are constructed as a two-component mixture of a point mass at zero and a sampling distribution for the count data (e.g., Poisson or negative binomial distributions in the univariate setting) \citep{xu2015assessment, zhang2020nbzimm,  jiang2021bayesian,shuler2021bayesian}. A corresponding latent indicator is introduced to differentiate between ``structural'' zeros which occur for events that have zero probability and ``at-risk'' zeros which occur for events that have positive probability but a zero count is still observed. Covariates may affect the sampling distribution of the counts as well as the probability of observing an at-risk observation \citep{neelon2019bayesian}. In multivariate settings, researchers link, or jointly model, zero-inflated univariate count models via latent parameters which govern the dependence structure between counts \citep{aitchison1989multivariate, chiquet2021poisson}. These approaches model multivariate counts unconditionally on the total count of a sample and are not fit for settings in which the count probabilities are defined on the simplex. As such, they are not suitable for the compositional count data collected in high-throughput sequencing settings when the total number of reads, or read depth, is fixed  \citep{gloor2017microbiome}.


%
\textcolor{black}{There are few methods available for modeling zero-inflated multivariate \textit{compositional} count data. Existing methods are limited as they make restrictive assumptions, fail to estimate parameter uncertainty, do not explicitly model zero-inflation indicators, only provide individual-level inference, and/or ignore potential covariates.}

Recently, \cite{tuyl2018method} proposed leveraging the neutrality of the Dirichlet distribution to allow compositional elements which have zero counts to potentially take on zero probabilities of occurrence. The task of determining which zero count observations are structural zeros is then cast as a model selection problem. The resulting mixture model is shown to reduce shrinkage when estimating the probability of categories with positive counts in the presence of zero count categories.
A major limitation of this approach is that it only provides count probability estimates for a single observation and is therefore not designed to provide population-level inference given a sample of potentially heterogeneous observations.

\cite{tang2019zero} introduced a zero-inflated \textit{generalized} DM model to detect group-wise differential mean and dispersion levels of microbial composition. The generalized Dirichlet is a conjugate prior for the multinomial distribution and is constructed from mutually independent beta distributed variables. \cite{tang2019zero} leverage the construction's stick-breaking formulation to model excess zeros by replacing the beta distributed variables with \textit{zero-inflated} beta distributed variables. They take an expectation-maximization (EM) approach for estimation, which provides a fast, parallelizable optimization procedure but lacks intrinsic uncertainty estimation. Further, their method is not designed for multiple regression settings.  \cite{zhou2021transformation} similarly proposed zero-inflated DM and Dirichlet-tree multinomial (DTM) models for differential abundance analysis. Their methods rely on a data augmentation strategy that induces dependence on the ordering of the compositional elements, similar to the GDM \citep{wong1998generalized} but unlike typical DM and DTM models. 

\cite{zeng2022zero} recently proposed a zero-inflated probabilistic principal component analysis logistic normal multinomial (ZIPPCA-lnm) model. Their method imposes a low-rank structure on the compositional data that accounts for complex correlation structures among counts and can flexibly incorporate observed covariates. The authors take an empirical Bayes approach for estimation that approximates the likelihood via variational techniques and maximizes the resulting objective function to obtain parameter estimates. They consider a naive mean-field variational approximation which assumes independence among all latent factors and excess zero indicators. To further improve convergence and reduce computation time, the authors impose a hard threshold on the probability of a structural zero within their optimization routine. Subsequently, the method relies on model comparison or cross-validation techniques for selecting the threshold level for excessive zeros, the number of factors in the model, as well as which covariates to include in the model. As a result, their approach may underestimate model uncertainty. Additionally, the current \texttt{R} implementation of the method is only designed to adjust for one observed covariate, precluding its use in multiple regression settings.

\cite{ren2017bayesian} also incorporate dependence on latent factors among compositional counts by assuming a marginal, dependent Dirichlet process prior for each composition which is truncated at the total number of observed components. As a result, their approach is able to assign a zero probability of occurrence for zero count categories, but it does not explicitly model zero-inflation indicators to differentiate at-risk and structural zeros which limits inference. \cite{ren2020bayesian} extend the work of \cite{ren2017bayesian} for mixed effects regression models but the approach is only designed to handle small to medium sized covariate spaces as it also relies on model comparison techniques for model selection.


\textcolor{black}{  A major limitation of existing methods for modeling zero-inflated multivariate compositional count data in exploratory microbiome research settings is that they are unable to perform variable selection on covariates associated with relative abundances and the probability of being an at-risk zero. While not designed for zero-inflated multivariate compositional count data, there are numerous variable selection methods available to explore potential relations between a high-dimensional set of covariates and microbial abundances using DM regression modeling frameworks and others \citep{chen2013variable,wang2017dirichlet,  wadsworth2017integrative,koslovsky2020microbvs,koslovsky2020microbiome,osborne2022latent,miao2020scalable}. Recently, \cite{jiang2021bayesian} proposed a Bayesian zero-inflated negative binomial regression model, which is able to identify subsets of taxa that are differentially abundant among subgroups in addition to performing variable selection. While their approach does not accommodate the compositional structure of the microbial abundance data and is not designed to identify covariates associated with potential zero-inflation, it has shown promising variable selection performance for covariates associated with multivariate compositional count data using discrete spike-and-slab prior formulations.}

\textcolor{black}{In this work, we propose a novel Bayesian zero-inflated Dirichlet-multinomial (ZIDM) model. While fully Bayesian methods are often criticized and even avoided in high-dimensional settings due to their computational demand, we take special care to devise a scalable approach which can handle the large model spaces encountered in omics research without sacrificing interpretability or imposing limiting assumptions. Specifically, we reparameterize the Dirichlet distribution via its relation to a set of normalized independent gamma random variables. We then replace the gamma distributions with zero-inflated gamma distributions to accommodate excess zeros. Further, we incorporate covariate dependence to model heterogeneity in compositional proportions as well as the probability of observing a structural zero.  To increase the scalabilty of our model, we introduce sparsity-inducing priors for corresponding regression coefficients and leverage the P\'{o}lya-Gamma data augmentation technique for efficient sampling and interpretability \citep{polson2013bayesian}.  Additionally for posterior inference, we propose a novel Metropolis-Hastings update for potentially zero-inflated individual-level relative abundances that can accommodate changes in the dimension of the model space across Markov chain Monte Carlo (MCMC) iterations.   We demonstrate the  estimation and selection performance of our model in numerous simulation scenarios and apply our model to zero-inflated microbial abundance data collected in the COMBO study. Compared to existing methods, our approach achieved improved or comparable estimation and variable selection performance on simulated data and higher variable selection stability estimates in application. }

\section{Proposed Model}\label{model}

    We first introduce the data augmentation technique used for efficient sampling of
    Bayesian DM models and propose our solution for accommodating zero-inflation. We then extend the model to regression settings and further embed sparsity inducing priors for regression coefficients to handle high-dimensional compositional and covariate model spaces, equipping the model for both  confirmatory and exploratory research settings.

\subsection*{Augmented Dirichlet-Multinomial Model }

 Let $\boldmath{z}_i' = (z_{i1}, \dots, z_{iJ})$ represent a $J$-dimensional vector of observed multivariate counts  collected on the $i^{th}$ observation,  $i = 1,\dots,N$.  We assume the counts $z_i$ follow a multinomial distribution
\begin{equation}\label{eq:one}
z_i\sim \mbox{Multinomial}(\dot{z}_{i}|\psi_i),
\end{equation}
where $\dot{z}_{i} = \sum_{j=1}^J z_{ij}$ is fixed, and $\psi_i= (\psi_{i1}, \dots, \psi_{iJ})$ with  $ \psi_{ij} \geq 0$ and $\sum_{j=1}^J \psi_{ij} = 1$.
To account for overdispersion in the multivariate count data, a common approach is to assume the compositional probabilities $\psi_i\sim \mbox{Dirichlet}({\boldmath\gamma}_i)$
with the $J$-dimensional vector $\boldmath{\gamma_i} = (\gamma_{ij}>0, \forall j \in J)$. Since the Dirichlet is a conjugate prior for the multinomial distribution, the posterior for $\psi_i$ also follows a Dirichlet distribution, with posterior mean estimates  $E_{p(\psi_i|z_i)}[\psi_{ij}]= \frac{z_{ij} + \gamma_{j}}{ \dot{z}_{i} + \sum_{j=1}^{J} \gamma_{j} }$,  for all $j=1,\dots,J$. Thus, $E_{p(\psi_i|z_i)}[\psi_{ij}] > 0$, even if the true probability of occurrence for $z_{ij}$ is zero. This property of the Dirichlet distribution is central to the methodological contributions of this work. 

While the conjugacy of the Dirichlet prior for a multinomial distribution can be exploited to help reduce computational demand for posterior inference, this is typically only the case in trivial settings.
In practice, hierarchical DM modeling frameworks typically rely on sampling-based methods for inference which are computationally burdensome due to the compositional structure and high-dimensionality of the data and resulting parameter space.  Instead of working directly with the Dirichlet distribution, we impose a data augmentation strategy inspired by techniques used in Bayesian nonparametrics \citep{james2009posterior, argiento2015priori} and detailed in \cite{koslovsky2020bayesian}. The advantages of this approach are two-fold. First, it  reduces the computational demand of the resulting  MCMC  algorithm in DM regression settings. Second, it facilitates an opportunity to introduce a zero-inflation indicator for each compositional element that allows the model to differentiate between a structural and at-risk zero  by letting $\psi_{ij}$ potentially take on zero values. 

To implement the data augmentation approach, we first define latent variables $c_{ij}$ such that $\psi_{ij} = c_{ij}/T_i$ with $T_i = \sum_{j=1}^Jc_{ij}$ and reparameterize Equation (Eq.) \eqref{eq:one}  as
$z_i \sim \mbox{Multinomial}(\dot{z}_i| c_i/T_i),$
where $ c_i' = (c_{i1},\dots,c_{iJ})$ and $c_{ij} \sim \mbox{Gamma}(\gamma_{ij},1)$. We then introduce auxiliary parameters
$u_i|T_i \sim \mbox{Gamma}(\dot{z_i},T_i)$ for $i=1\dots,N$. \textcolor{black}{ This approach greatly reduces computational demand and improves mixing by eliminating unnecessary calculations of $T_i$ when sampling the posterior distribution and providing Gibbs updates for $c_{ij}$ and $u_i$. See the Supporting Information for more technical details.}

\subsection*{Zero-Inflated Dirichlet Distribution}
Leveraging the data augmentation technique presented in the previous section, we propose a zero-inflated Dirichlet distribution. Intuitively, we seek an approach that places a point mass at zero for $\psi_{ij} $ when $z_{ij} = 0$ represents a structural zero. Since $\psi_{ij} = c_{ij}/T_i$,  $\psi_{ij} = 0$ when $c_{ij} = 0$. Therefore to model potential zero-inflation, we introduce an at-risk indicator variable $\eta_{ij} \in \{0,1\}$ for all $i = 1,\dots, N$ and $j = 1, \dots, J$ where $\eta_{ij} = 0$ indicates a structural zero (i.e., $c_{ij} =0$) and $\eta_{ij} = 1$ indicates $c_{ij} > 0$. Specifically, we assume a zero-inflated Gamma distribution for  $c_{ij}$, $
    c_{ij}^{(\eta)}|\eta_{ij} \sim (1-\eta_{ij})\delta_0(\cdot) + \eta_{ij}\mbox{Gamma}(\gamma_{j},1)$,
where $\eta_{ij} \sim \mbox{Bernoulli}(\theta_j)$ and $\theta_j = \frac{\exp({\beta_{\theta j0}})}{1+\exp({\beta_{\theta j0}})}$ ($\Theta_j = 1-\theta_j$) is the probability of a non-zero $c_{ij}$ (structural zero) for the $j^{th}$ compositional element. The superscript $(\eta)$ reflects the dependence of $c_i$ and subsequently the sampling distribution of $z_i$ on the at-risk indicator. By assigning zero values to a subset of the compositional probabilities, the dimension of $c_i^{(\eta)}$ and corresponding $z_i$ is reduced to $\sum_{j=1}^{J} \eta_{ij}.$ Note that when $\eta_{ij} = 1$, $z_{ij}$ may still equal  zero, but when $\eta_{ij} = 0$, $z_{ij} = 0$.
 
\subsection*{ZIDM Regression Model}

We present a general framework for the proposed ZIDM   model in which count probabilities and at-risk indicators depend on covariates. To generate inference on the relation between each compositional element and each covariate, we set $\lambda_{ij}  = \log(\gamma_{ij} )$ and assume $
  \lambda_{ij}  =  x_i^{\prime}\boldsymbol{\beta}_{\gamma j},$
where $x_i^{\prime} = (1, x_{i1},\dots, x_{i,P-1})$ represents a $P$-dimensional vector of covariates for the $i^{th}$ observation including an intercept term and  $\boldsymbol{\beta}_{\gamma j}^{\prime}  = (\beta_{\gamma j0} ,\beta_{\gamma j1} ,\dots,\beta_{\gamma j,P-1} )$ represents the corresponding covariates' relation with the $j^{th}$ compositional element. By exponentiating $\lambda_{ij}$, we ensure positive hyperparameters for the zero-inflated Dirichlet distribution. Here, the exponentiation of a regression coefficient is interpreted as the multiplicative factor of change in the proportion of a compositional element with a one unit change in the corresponding standardized covariate while holding all else constant \citep{chen2013variable}.

\subsection*{Sparsity-Inducing Priors}
For high-dimensional covariate spaces, we propose embedding multivariate variable selection spike-and-slab priors for $\boldsymbol{\beta}_{\gamma j} $ to encourage sparsity in the relation between covariates and the multivariate count data, similar to \cite{wadsworth2017integrative,koslovsky2020bayesian,koslovsky2020microbvs,osborne2022latent}. We assume the covariates' inclusion in the model is characterized by a latent $J \times P$-dimensional inclusion vector $\boldsymbol{\varphi} $. With this formulation, $\varphi_{jp} =1$ indicates that covariate $p$ is associated with compositional element $j$ and 0 otherwise. The prior for $\beta_{\gamma jp} $ given $\varphi_{jp}$ follows a mixture of a normal distribution and a Dirac-delta function at zero, $\delta_{0}$, and is commonly referred to as the  spike-and-slab prior \citep{george1997approaches,brown1998multivariate}. Specifically,
$ \beta_{\gamma jp} |\varphi_{jp} ,\sigma_{\beta_{\gamma}}^2 \sim \varphi_{jp}  \cdot N(0,\sigma_{\beta_{\gamma}}^2) + (1-\varphi_{jp} ) \cdot \delta_0(\beta_{\gamma jp} ),
$
where $\sigma_{\beta_{\gamma}}^2$ is a diffuse variance. We assume each $\varphi_{jp} $ follows a Bernoulli prior, $p(\varphi_{jp} ) \sim \mbox{Bernoulli}(w_{jp} )$,
where $w_{jp}  \sim \mbox{Beta}(a_\varphi,b_\varphi)$. Integrating out $w_{jp}$ leads to 
$
p(\varphi_{jp} ) =  \mbox{Beta}(\varphi_{jp} + a_{\varphi},1 - \varphi_{jp} + b_{\varphi})/\mbox{Beta}(a_{\varphi},b_{\varphi}).$ Hyperparameters $a_{\varphi}$ and $b_{\varphi}$ can be set to impose various levels of sparsity in the model. Note that covariates, including the intercept term, can be forced into the model by fixing $\varphi_{jp} = 1$ for implementation of a standard ZIDM regression model. 
 
    Additionally, we allow the probability of $c_{ij}>0$ to depend on an observed set of covariates by replacing  $\theta_j$ with  $ \theta_{ij} =  \exp(x_i^\prime\boldsymbol{\beta}_{\theta j})/(1+ \exp(x_i^\prime \boldsymbol{\beta}_{\theta j}))$, where $\boldsymbol{\beta}_{\theta j}^{\prime} = (\beta_{\theta j0},\dots, \beta_{\theta j,P-1})$. We incorporate a latent inclusion indicator $\zeta_{jp}$ for $\beta_{\theta jp}$ to induce sparsity in the covariates associated with at-risk observations. Specifically, we assume $\beta_{\theta jp}|\zeta_{jp}, \sigma^2_{\beta_{\theta}} \sim \zeta_{jp} \cdot N(0,\sigma_{\beta_{\theta}}^2) + (1-\zeta_{jp})\cdot\delta_0(\beta_{\theta jp})$ with $p(\zeta_{jp}) =  \mbox{Beta}(\zeta_{jp} + a_{\zeta},1 - \zeta_{jp} + b_{\zeta})/\mbox{Beta}(a_{\zeta},b_{\zeta})$. Again, setting $\zeta_{jp} = 1$ forces the corresponding covariate into the model. Note that the covariate set potentially associated with count probabilities may differ from those potentially associated with excessive zeros.


\subsection{Posterior Inference}\label{posterior}

\begin{figure}[ht] 
\centering
          \includegraphics[scale=0.45]{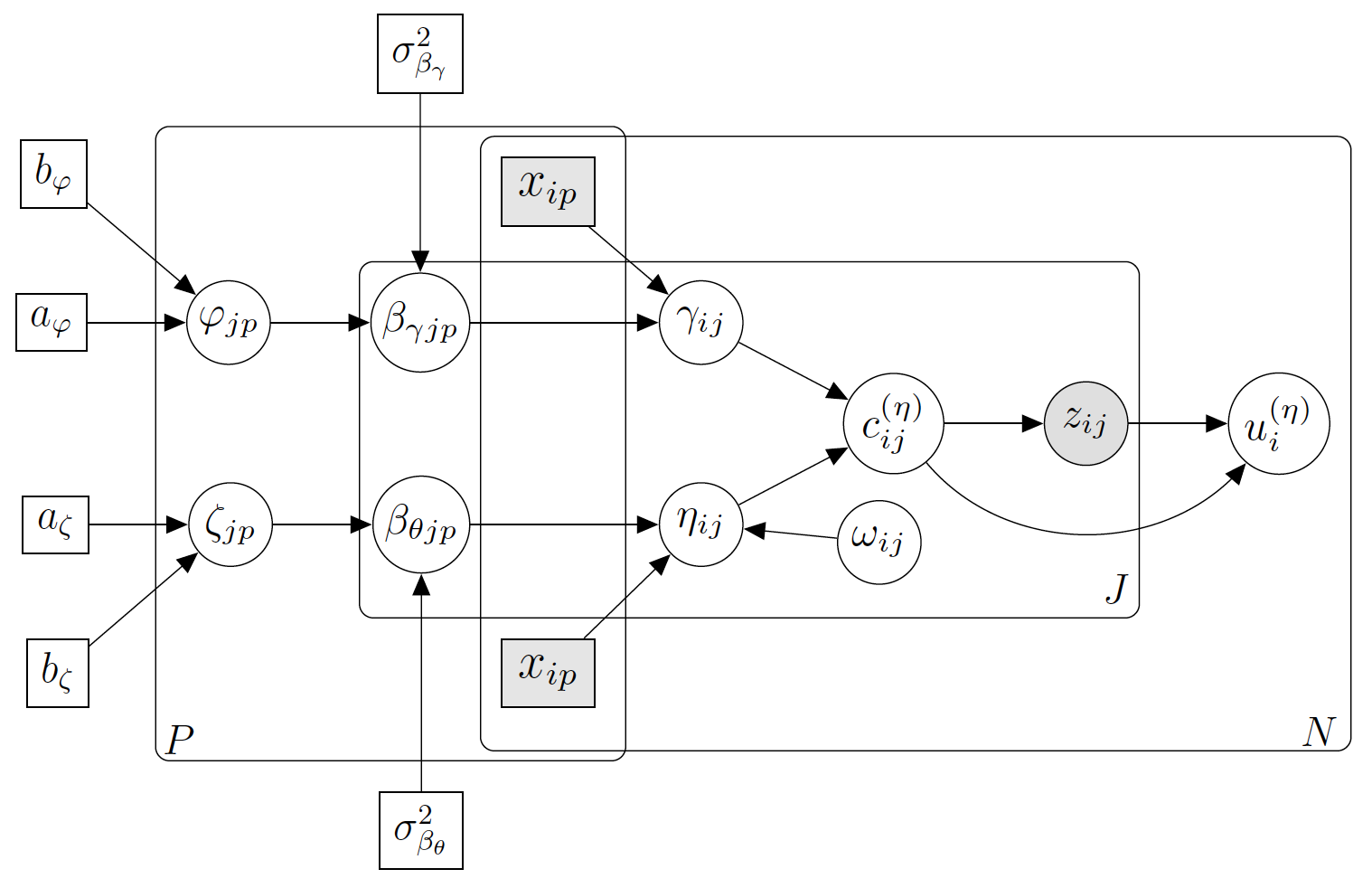}
  \caption{ \textcolor{black}{ Graphical representation of the proposed ZIDM regression model with sparsity-inducing priors at both levels of the model. $\beta_{\gamma jp}$ ($\varphi_{jp}$) and $\beta_{\theta jp}$ ($\zeta_{jp}$) represent the regression coefficients (corresponding latent inclusion indicators) for covariates ($x_{ip}$) associated with observed microbial abundances ($z_{ij}$) and zero-inflation probabilities, respectively. With no covariates in the model, $\Theta_j = 1/(1+\exp(\beta_{\theta j0}))$ represents the population-level zero-inflation probabilities, $\Gamma_j = \exp(\beta_{\gamma j0})/ (\sum_{j=1}^J\exp(\beta_{\gamma j0}))$ the population-level count probabilities, and $\psi_{ij} = c_{ij}^{(\eta)}/T_i^{(\eta)}$ the individual-level count probabilities, with $T_i^{(\eta)}  = \sum_{j=1}^Jc_{ij}^{(\eta)}$ for all $i=1,\dots,N$ and $j=1,\dots,J$, respectively. Circular (square) nodes represent random (fixed) variables and shaded (white) nodes represent observed data (parameters).  Plates denote replication. $N$ - total observations; $J$ - total compositional elements; $P$ - number of covariates including the intercept term. }    }\label{dag}
\end{figure}

For posterior inference, we construct a Metropolis-Hastings within Gibbs sampler.  The full joint  distribution is defined as  $$\prod_{i = 1}^N f(z_i|c_i^{(\eta)})p(u_i^{(\eta)}|z_i,c_i^{(\eta)}) \prod_{j=1}^J f(c_{ij}^{(\eta)}|x_i, \boldsymbol{\beta}_{\gamma j})p( \eta_{ij}|\omega_{ij},x_i,\boldsymbol{\beta}_{\theta j}) p( \omega_{ij})  p(\boldsymbol{\beta}_{\gamma j}|\boldsymbol{\varphi}_j) p(\boldsymbol{\varphi}_j) p(\boldsymbol{\beta}_{\theta j}|\boldsymbol{\zeta}_j)
   p(\boldsymbol{\zeta}_j), $$  
where auxiliary parameters $\omega_{ij}$ are introduced for each $\eta_{ij}$ to provide efficient sampling and interpretability of $\boldsymbol{\beta}_{\theta j}$ using the data augmentation technique of \cite{polson2013bayesian}. A graphical representation of the proposed model is presented in Figure (Fig.) \ref{dag}.  The MCMC sampler used to implement our model is outlined below in Algorithm \ref{MCMC}. 
\begin{algorithm} 
	\caption{MCMC Sampler}\label{MCMC}
	\footnotesize
	\begin{algorithmic} 
		\State Input data $z_i$ and $x_i$ for all $i = 1,\dots,N$.
		\State Initialize parameters: $c_i$, 
		$u_i$, $\boldsymbol{\beta}_{\theta j}$, $\boldsymbol{\beta}_{\gamma j}$, $\boldsymbol{\zeta}_j$, $\boldsymbol{\varphi}_j$, for all $i=1,\dots,N$ and $j=1,\dots,J$, respectively.
		
		\State Specify hyperparameters: $ \sigma^2_{\beta_\theta}, \sigma^2_{\beta_\gamma} a_{\varphi},b_{\varphi},a_{\zeta},b_{\zeta}$.
		\For{iteration $m = 1,\dots,M$}
		 \For{$i = 1,\dots,N$}
		 \State Update $u_{i}^{(\eta)} \sim \mbox{Gamma}(\dot{z}_i, T_{i}^{(\eta)})$. 
		\For{ $j = 1,\dots,J$}
		\State Update $\omega_{ij} \sim \mbox{PG}(1, \tau_{ij})$, where $\tau_{ij} = x_{i}^{\prime} \beta_{\theta j}$   via \cite{polson2013bayesian}.
		\State Jointly update $c_{ij}^{(\eta)}$ and $\eta_{ij}$ with an Expand/Contract Step.
		\State Update $c_{ij}^{(\eta)} \sim \mbox{Gamma}(z_{ij} + \gamma_{ij}, 1 + u_i)$.
		\EndFor
	    \EndFor
		 
		\State Jointly update $\boldsymbol{\beta}_{\gamma}$ and $\boldsymbol{\varphi}$ with Between and Within Steps via \cite{savitsky2011variable}. 
		\State Jointly update $\boldsymbol{\beta}_{\theta}$ and $\boldsymbol{\zeta}$ with Between and Within Steps via \cite{savitsky2011variable}. 
	    \EndFor
	\end{algorithmic}
\end{algorithm}
While similar to the two-component zero-inflated mixture models designed for univariate count data, our approach for multivariate compositional count data differs in that we assume a mixture distribution on the count \textit{probabilities} as opposed to the sampling distribution of the counts. As a result, the dimension of the parameter space changes as the MCMC algorithm iterates through various combinations of at-risk observations and structural zeros. Specifically, the dimension of the count probabilities $ \psi_i$ grows or shrinks as $\eta_{ij}$ transitions from $1$ to $0$ or $0 $ to $1$ iteration-to-iteration. To address this, we propose jointly updating $\eta_{ij}$ and $c_{ij}$ in what we refer to as an Expand or Contract Step (see Supporting Information for details). Note that our approach is reminiscent of the sampler proposed by \cite{savitsky2011variable} to traverse a regression coefficient space whose complexity changes over MCMC iterations. \textcolor{black}{Details of the  MCMC algorithm and model identifiability  are found in the Supporting Information. }

 After burn-in, the remaining samples obtained from running Algorithm 1 for $M$ iterations are used for inference. To identify covariates associated with count probabilities and at-risk observations, their corresponding marginal posterior probabilities of inclusion (MPPIs) are empirically estimated by calculating the average of their respective inclusion indicator's MCMC samples \citep{george1997approaches}. Typically, covariates are included in the model if their MPPI exceeds 0.50 \citep{barbieri2004optimal} or a Bayesian false discovery rate threshold, which controls for multiplicity \citep{newton2004detecting}.

\textcolor{black}{The per-iteration time
and space complexity of the proposed Metropolis-Hastings within Gibbs algorithm are linear with sample size $N$ but depend greatly on the sparsity at both levels of the model. For large compositional spaces, the overall time (space) complexity of each MCMC iteration is dominated by the Expand and Contract Step (Within Step for $\beta_{\theta jp}$), $\mathcal{O}(J^2NP)$ ($\mathcal{O}(NP+NJ+JP+P^2)$). One of the benefits of the discrete spike-and-slab prior is that for sparse models (i.e., $\sum_{p=1}^P \zeta_{jp} << P$ and/or $\sum_{p=1}^P \varphi_{jp} << P$  ), the time and space complexities are greatly reduced.  See the Supporting Information for details of the model's computational complexity calculations.}

\section{Simulations}\label{simulations}

In this section, we evaluate and compare the performance of the proposed ZIDM model using simulated data in three scenarios with various data generation settings. The first scenario examines the estimation performance of the ZIDM model with respect to population-level zero-inflation probabilities, population-level count probabilities, and individual-level count probabilities. Using the proposed model's notation, these quantities are $\Theta_j = 1/(1+\exp(\beta_{\theta j0}))$, $\Gamma_j = \exp(\beta_{\gamma j0})/ (
\sum_{j=1}^J\exp(\beta_{\gamma j0}))$, and $\psi_{ij} = c_{ij}/T_i$ for all $i=1,\dots,N$ and $j=1,\dots,J$, respectively. Note that in this scenario, the ZIDM model only estimates an intercept term in both levels of the model and ignores any potential covariates.  We compare the ZIDM model to a Bayesian DM model, Tuyl's approach \citep{tuyl2018method}, ZIPPCA-lnm \citep{zeng2022zero}, and DirFactor \citep{ren2017bayesian} when applicable. 

In the second and third scenarios, we incorporate covariates into both levels of the model and investigate  variable selection performance, in addition to individual-level estimation of the zero-inflation probabilities, $\Theta_{ij}$, and $\psi_{ij}$. To our knowledge, there are no other existing methods that perform variable selection in zero-inflated multivariate compositional regression models for direct comparison. \textcolor{black}{Thus, we compare our model's variable selection performance to a DM regression model with spike-and-slab priors (DMbvs) presented in \cite{wadsworth2017integrative},  the penalized DM approach of \cite{chen2013variable} (DMpen), as well as a Bayesian variable selection method for zero-inflated negative binomial regression models recently proposed by \cite{jiang2021bayesian}  (ZINB)}. For clarity, we denote our proposed method as ZIDMbvs when it is used for variable selection. For comparison, we implemented the Bayesian DM model, DMbvs, and Tuyl's approach in \texttt{Rcpp} \citep{eddelbuettel2011rcpp}, similar to our methods (i.e., ZIDM and ZIDMbvs). \textcolor{black}{Implementation of ZIPPCA-lnm, DirFactor, and  ZINB is achieved via their corresponding \texttt{R} packages, ZIPPCA-lnm, DirFactor-fix, and IntegrativeBayes, \nocite{ZIPPCAlnm} \nocite{DirFactor}   \nocite{IntegrativeBayes} respectively.}

In each scenario, we simulated various numbers of individuals, $N$, compositional components, $J$, and covariates, $P$. \textcolor{black}{ Multivariate count data were sampled from a Multinomial($\dot{z}_i| \psi_i^*$), where the total number of counts $\dot{z}_i$ was simulated from a uniform distribution with varying upper and lower bounds to induce different levels of zero counts. The individual-specific count probabilities $\psi_i^*$  were assumed to follow a $ \mbox{Dirichlet}(\gamma_i^*)$, where $\gamma_i^* = (\gamma_{i1}^*,\gamma_{i2}^*,\dots,\gamma_{iJ}^*)$. Each $\gamma_{ij}^* = \frac{\gamma_{ij}*\eta_{ij}}{\sum_{j=1}^J \gamma_{ij}*\eta_{ij}}\frac{1-d}{d}$, $j = 1,\dots,J$, where $\gamma_{ij}$ is defined above, $\eta_{ij} \sim \mbox{Bernoulli}(\theta_{ij})$, and $d$ serves as an overdispersion parameter which was set at $0.01$, similar to \cite{wadsworth2017integrative}. Thus, the data generating model differs from all  methods compared in this study.} Covariates used to define $\gamma_{ij}$ and $\theta_{ij}$ were simulated from a $N_{P-1}(\boldsymbol{0},\Sigma)$, where $\Sigma_{st} = \sigma^{|s-t|}$.

Each of the Bayesian methods were run for 20,000 iterations and thinned to every $10^{th}$ iteration. This resulted in 2,000 iterations, of which the first 1,000 iterations were treated as burn-in, and the remaining 1,000 used for inference. We assumed weakly-informative diffuse variances $\sigma^2_{\beta_\gamma} = 10$ and $\sigma^2_{\beta_\theta} = 10$ for regression coefficients, unless otherwise specified. We assumed a non-informative prior probability of inclusion at both levels of the model with $a_\varphi = a_\zeta = b_\varphi = b_\zeta = 1$ when necessary, and the intercept terms were forced into the model by fixing their latent inclusion indicators to one. The spike-and-slab prior specifications for ZINB were set similar to the other Bayesian methods for consistency. All regression coefficients were initiated at zero, with the exception of the intercept terms $ \beta_{\gamma j0} $ and $ \beta_{\theta j0} $, which were simulated from a standard normal. We initialized $\eta_{ij}|z_{ij} \neq 0 \sim \mbox{Bernoulli}(0.5)$ and $\omega_{ij} = 1$. The ZIPPCA-lnm and DirFactor models were run with default settings. Since the true number of factors is unknown, we fit the ZIPPCA-lnm  model with 1 to 5 factors and report the results from the model with the lowest Bayesian information criterion, as recommended by \cite{zeng2022zero}. The \texttt{ZIPPCAlnm} package provides 95\% confidence intervals for $\Theta_j$ and latent factors using a sandwich estimator but does not provide direct uncertainty estimates for $\Gamma_j$ or $\psi_{ij}$. DirFactor only provides point and uncertainty estimates for individual-level count probabilities, $\psi_{ij}$. For Tuyl's approach, we obtained 95\% confidence intervals using Monte Carlo sampling with 40,000 iterations.  \textcolor{black}{ Note that DMbvs, DMpen, and ZINB only perform variable selection for covariates potentially associated with the compositional relative abundances.}

To evaluate the estimation performance of the models, we calculated the average absolute value of the difference between the estimated and true probabilities (ABS), Frobenius norm, FROB = $\sqrt{\sum_{i = 1}^N\sum_{j=1}^p(\hat{\rho}_{ij} - \rho_{0ij})^2}$, Simpson's index mean squared error, SIMP = $1/N \sum_{i=1}^N(\sum_{j=1}^J\rho^2_{0ij} - \sum_{j=1}^J \hat{\rho}^2_{ij})^2$, and  95\% coverage probabilities (COV), where $\hat{\rho}_{ij}$ and $\rho_{0ij}$, represent estimated and true probabilities, respectively. We adjusted these metrics for population-level parameters as necessary. For variable selection, the methods were assessed on the basis of sensitivity (1 - false negative rate), specificity (1 - false positive rate), Matthew's correlation coefficient (MCC), and F1 score (two measures of overall selection accuracy). These are defined as 
$\mbox{Sensitivity} = \frac{TP}{FN + TP}$,
$\mbox{Specificity} = \frac{TN}{FP + TN}$,
$MCC = \frac{TP \times TN - FP \times FN}{\sqrt{(TP +FP)(TP+FN)(TN+FP)(TN+FN)}} ,$
$F1 = \frac{2TP }{2TP + FP + FN} ,$
where TN, TP, FN, and FP represent the true negatives, true positives, false negatives, and false positives, respectively. Additionally, we compare the computation time of each method run on an Intel Xeon Bronze 3204 1.9 GHz processor with 16 GB RAM.   Results we report below were obtained by averaging over 100 replicated data sets for each setting. 

\subsection{Scenario 1 }

In this section, we evaluate and compare the estimation performance of the proposed method. We first set $N=100$ and $J=50$, where $\dot{z}_i \sim \mbox{Uniform}(400,500)$ with baseline zero-inflation parameters, $\beta_{\theta j0}$, set to range between 0 and 1, inducing 50\% zeros of which 25\% were at-risk on average. Baseline compositional count parameters $\beta_{\gamma j0}$ were randomly sampled from Uniform(-2.3, 2.3). Results of this setting are presented in Table \ref{tab:setting 1}. Overall, we found that ZIDM better estimated the population-level zero-inflation probabilities $\Theta_j$ and count probabilities $\Gamma_j$ compared to the ZIPPCA-lnm and the DM  models, respectively. Note that the DM (ZIPPCA-lnm) model does not provide estimates for $\Theta_j$ ($\Gamma_j$). All five methods performed relatively well estimating the individual-level count probabilities $\psi_{ij}$, with DirFactor and Tuyl's approach demonstrating a slight advantage. Note that neither of these methods provide estimates for $\Theta_j$ and $\Gamma_j$.  Our approach was able to obtain nominal coverage probabilities for $\Theta_j$ and $\Gamma_j$, but all methods obtained roughly a 30\% coverage probability for $\psi_{ij}$ (excluding ZIPPCA-lnm which does not estimate individual-level count probability uncertainty). The DM, ZIPPCA-lnm, and ZIDM methods took roughly 1, 2, and 4 minutes to run, respectively. Tuyl's approach provided point estimates in less than a second but required around 7 minutes to generate the Monte Carlo samples for uncertainty estimation. DirFactor took over 25 minutes to generate the 20,000 MCMC samples.

\begin{table}[]
\caption{ Simulation Results: Parameter estimation performance in Scenario 1 for $N=100$ observations and $J=50$ compositional components with 50\% zero cells of which 25\% are at-risk on average. ABS - absolute value of the difference between the estimated and true probabilities; FROB - Frobenius norm; SIMP - Simpson's index mean squared error; COV  - 95\% coverage probabilities. Time is in seconds (s). Time for Tuyl's approach refers to point estimate runtime with Monte Carlo sampling runtime in parentheses.   }

\label{tab:setting 1}
\setlength{\tabcolsep}{2.5pt}  
\centering 
\begin{tabular}{ccccccc}
\hline
   Model        & Parameter                       & ABS   & FROB  & SIMP & COV & Time (s)     \\ \hline
ZIDM       & \multirow{2}{*}{$\Theta_j$} & 0.068 & 0.636 & 0.447     & 0.953    & -            \\
ZIPPCA-lnm &                                 & 0.130 & 1.245 & 44.995   & 1.000    & -            \\ \hline
ZIDM       & \multirow{2}{*}{$\Gamma_j$} & 0.001 &  0.013     &    3.393$\times10^{-6}$      & 0.965    & -            \\
 DM         &                                 & 0.006 &  0.060     &   0.001        & 0.003    & -            \\ \hline
ZIDM       & \multirow{4}{*}{$\psi_{ij}$}    & 0.014 & 1.842 & 2.493$\times10^{-4}$ & 0.252    & 233.3        \\
 DM         &                                 & 0.015 & 1.912 & 2.756$\times10^{-4}$ & 0.227    & 60.7         \\
Tuyl's     &                                 & 0.008 & 1.017 & 1.241$\times10^{-4}$ & 0.343    & 0.9 (410.0) \\
ZIPPCA-lnm &                                 & 0.014 & 1.616 & 2.724$\times10^{-3}$ & -    & 134.4        \\
DirFactor &                                 & 0.007 & 1.015 & 1.820$\times10^{-4}$ & 0.344    & 1634.2        \\ \hline
\end{tabular}
\end{table}

 We further examined the estimation performance of the model in a variety of settings which are detailed in the Supporting Information. Briefly, we explored the models' estimation performance with varying levels of at-risk zeros and sample sizes, as well as with different data generation processes including \textcolor{black}{under the assumption of the ZIPPCA-lnm model and a negative multinomial distribution}. Overall, the relative performance of the methods was quite similar as in the above settings, and the proposed model was fairly robust to model misspecification. \textcolor{black}{For larger sample sizes (i.e., $\geq 500$ compositional components), Tuyl's approach and ZIPPCA-lnm often failed to provide results, due to memory constraints, numerical issues, and/or failed convergence. However, results were comparable to the proposed method, DM, and DirFactor, when available. }

 \subsection{Scenario 2 }\label{scenario2}

In the second simulation scenario, the total number of counts, $\dot{z}_i$, were simulated from a \mbox{Uniform}$(1000,2000)$ with baseline zero-inflation parameters, $\beta_{\theta j0}$, and compositional count parameters, $\beta_{\gamma j0}$, set similar to Scenario 1. In each of the 100 replicate data sets, we set 16 of the $J*(P-1)$ regression coefficients to be active in both levels of the model (in addition to the intercept terms). Corresponding regression coefficients were randomly sampled from  $ \pm[0.9,1.5]$, and the  covariates' correlation was specified with $\sigma = 0.3$. 

\begin{table}[]
\caption{ Simulation Results: Variable selection performance in Scenario 2  for covariates in the zero-inflation and DM portions of the model with corresponding coefficients $\boldsymbol{\beta}_\theta$ and $\boldsymbol{\beta}_\gamma$, respectively. SENS - sensitivity;  SPEC - specificity; MCC - Matthew's correlation coefficient; F1 - F1 score.  \label{tab:setting2selection}}
\begin{tabular}{ccccccccc}
\hline
N                    & J                    & P                    & Model   & Coefficients                    & SENS                 & SPEC                 & MCC                  & F1                   \\ \hline
 50   &  100  &  50 & ZIDMbvs & $\boldsymbol{\beta}_\theta$     & 0.452               & 0.962               & 0.153                & 0.106               \\ \cline{4-9} 
                     &                      &                      & ZIDMbvs & \multirow{4}{*}{$\boldsymbol{\beta}_\gamma$} & 0.766                & 0.979               & 0.358                & 0.284               \\
                     &                      &                      & DMbvs   &                                 & 0.332                & 0.980                & 0.178                & 0.159               \\
  &  &   & \textcolor{black}{DMpen}    & & \textcolor{black}{0.137} & \textcolor{black}{0.984}   & \textcolor{black}{0.055}  & \textcolor{black}{0.047}  \\
  &  &   & \textcolor{black}{ZINB}    & &  \textcolor{black}{0.613} &  \textcolor{black}{0.980} & \textcolor{black}{0.233}  & \textcolor{black}{0.161} \\ \hline
100                  & 50                   & 100                  & ZIDMbvs & $\boldsymbol{\beta}_\theta$     & 0.808               & 0.953               & 0.275               & 0.180              \\ \cline{4-9} 
                     &                      &                      & ZIDMbvs & \multirow{4}{*}{$\boldsymbol{\beta}_\gamma$} & 0.955               & 0.983                & 0.509                & 0.428               \\
                     &                      &                      & DMbvs   &                                 & 0.147                & 0.996               & 0.175                & 0.030               \\
  &  &  & \textcolor{black}{DMpen} & & \textcolor{black}{0.408} & \textcolor{black}{0.977} & \textcolor{black}{0.195} & \textcolor{black}{0.164} \\
  &  &  & \textcolor{black}{ZINB} & & \textcolor{black}{0.945} & \textcolor{black}{0.981} & \textcolor{black}{0.482} & \textcolor{black}{0.398} \\ \hline
100                  & 500                  & 50                   & ZIDMbvs & $\boldsymbol{\beta}_\theta$     & 0.534              & 0.894            & 0.045               & 0.011             \\ \cline{4-9} 
                     &                      &                      & ZIDMbvs & \multirow{4}{*}{$\boldsymbol{\beta}_\gamma$} & 0.797              & 0.964              & 0.132               & 0.011              \\
                     &                      &                      & DMbvs   &                                 & 0.706             & 0.925             & 0.014               & 0.044               \\
  &  &  & \textcolor{black}{ZINB} & & \textcolor{black}{0.727} & \textcolor{black}{0.865} & \textcolor{black}{0.086} & \textcolor{black}{0.023} \\ \hline
\end{tabular}
\end{table}

The selection performance results of our simulation study with varying numbers of observations, compositional components, and covariates are presented in Table \ref{tab:setting2selection}. \textcolor{black}{Note that only the proposed method is able to perform variable selection on covariates potentially associated with the at-risk indicators.} Here, we observed relatively stable specificity levels across simulation settings, but the sensitivity of the proposed method improved with increased sample size, as expected. \textcolor{black}{We found that ZIDMbvs outperformed DMbvs, DMpen, and ZINB in terms of selection performance for covariates associated with the count data in the presence of zero-inflation. In results not shown, we found that the selection performance was similar with larger covariate spaces. Additionally when data were simulated without zero-inflation, we found that our proposed model maintained similar performance to DMbvs and outperformed DMpen and ZINB in terms of sensitivity and specificity  (Web Table S1).  For large $J$ settings, we found that DMpen and ZINB often failed to provide results due to due to memory constraints, numerical issues, and/or failed convergence.}


 \subsection{ \textcolor{black}{Scenario 3} }\label{scenario3}
\textcolor{black}{In the third simulation scenario, we generated data similar to the application data (Section \ref{application}) with the total number of counts, $\dot{z}_i$, simulated from a \mbox{Uniform}$(1100,15000)$  and baseline zero-inflation parameters, $\beta_{\theta j0}$, set to induce varying levels of zero-inflation (including a scenario with 30\% zeros, similar to the application data). The compositional count parameters, $\beta_{\gamma j0}$, were set similar to Scenario 1. In each of the 100 replicate data sets, we set 16 of the $J*(P-1)$ regression coefficients to be active in both levels of the model (in addition to the intercept terms). Corresponding regression coefficients were sampled from  $ \pm[1.0,3.0]$, and the covariates' correlation was specified with $\sigma = 0.8$. }

\textcolor{black}{The selection performance results of our simulation study with data generated similar to the structure of the application data are presented in Table \ref{tab:setting3selection}.  We found that the overall variable selection performance (MCC and F1) of the proposed method remained consistent across varying levels of zeros and at-risk zeros and performed the best overall. While the ZINB method often obtained higher sensitivity than the proposed method in this setting, it underperformed with respect to specificity. We also observed that the sensitivity for all methods except DMpen increased as the number of zeros in the data decreased. 
} 

\textcolor{black}{Additionally to assess the models' selection performance with misspecification, we generated data from a negative multinomial distribution with varying levels of random noise introduced for the covariate dependent count probabilities. We observed similar results as Scenario 3, in which ZIDMbvs obtained the best performance overall, but ZINB had the highest sensitivity. See the Supporting Information for details.}
 
 \begin{table}[]
    \caption{ Simulation Results: Variable selection performance in Scenario 3 with varying levels of zeros and at-risk zeros and for covariates in the zero-inflation and DM portions of the model with corresponding coefficients $\boldsymbol{\beta}_\theta$ and $\boldsymbol{\beta}_\gamma$, respectively.  SENS - sensitivity; SPEC - specificity; MCC - Matthew's correlation coefficient; F1 - F1 score.\label{tab:setting3selection}}
\begin{tabular}{cccccccc}
\hline
\% zeros & \multicolumn{1}{l}{\% at-risk} & Model   & Coefficients                    & SENS   & SPEC   & MCC                        & F1                         \\ \hline

50       & 25                             & ZIDMbvs & $\boldsymbol{\beta}_\theta$     & 0.698 & 0.970 & 0.336                    & 0.316                     \\ \cline{3-8} 
         &                                & ZIDMbvs & \multirow{4}{*}{$\boldsymbol{\beta}_\gamma$} & 0.794 & 0.959 & 0.376                     & 0.307                     \\
         &                                & DMbvs   &                                 & 0.383 & 0.973 & 0.223                     & 0.208                     \\
         &                                & DMpen     &                                 & 0.886 & 0.396 & 0.065                     & 0.899                     \\
         &                                & ZINB      &                                 & 0.920 & 0.910 & 0.303                    & 0.201                     \\ \hline
40       & 50                             & ZIDMbvs & $\boldsymbol{\beta}_\theta$     & 0.638 & 0.978 & 0.386                     & 0.354                    \\ \cline{3-8} 
         &                                & ZIDMbvs & \multirow{4}{*}{$\boldsymbol{\beta}_\gamma$} & 0.844 & 0.964 & 0.443                     & 0.352                     \\
         &                                & DMbvs   &                                 & 0.521 & 0.967 & 0.277                     & 0.246                     \\
         &                                & DMpen      &                                 & 0.736 & 0.709 & 0.208                     & 0.170                     \\
         &                                & ZINB      &                                 & 0.934 & 0.922 & 0.331                     & 0.228                     \\ \hline
30      & 70                             & ZIDMbvs & $\boldsymbol{\beta}_\theta$     & 0.506 & 0.984 & 0.355                     & 0.344                     \\ \cline{3-8} 
         &                                & ZIDMbvs & \multirow{4}{*}{$\boldsymbol{\beta}_\gamma$} & 0.856 & 0.961 & 0.410                     & 0.334                     \\
         &                                & DMbvs   &                                 & 0.598 & 0.966 & 0.316                     & 0.275                     \\
         &                                & DMpen      &                                 & 0.719 & 0.796 & \multicolumn{1}{l}{0.248} & \multicolumn{1}{l}{0.205} \\
         &                                & ZINB      &                                 & 0.954 & 0.921 & 0.330  & 0.223   \\ \hline
\end{tabular}
\end{table}

\subsection{Sensitivity Analysis}\label{sec::sens}
 In this section, we investigate ZIDMbvs's sensitivity to specification of hyperparameters $\sigma^2_{\beta_\theta}$,  $\sigma^2_{\beta_\gamma}$, $a_\varphi$, $a_\zeta$, $b_\varphi$, and $b_\zeta$. In each of the sensitivity analyses, replicate data were generated from the
model defined in Section \ref{scenario2}. To assess the model's sensitivity to hyperparameter settings, we set each of the hyperparameters to default values and then evaluated the effect of manipulating each term on parameter estimation and selection performance. For the default parameterization, we set the hyperparameters  $\sigma^2_{\beta_\theta} = 10$,  $\sigma^2_{\beta_\gamma} =10$, and $a_\varphi = a_\zeta = b_\varphi = b_\zeta = 1$. A sensitivity analysis of DMbvs with similar parameterizations is provided for comparison.

The results of the sensitivity analysis are presented in Web Tables S2 and S3. We observed very little sensitivity in terms of individual-level zero-inflation and count probability estimation with more sparsity induced in the model a priori. The estimation performance was also unaffected by the assumed variances. As expected, the number of selected covariates in both levels of the model decreased as the prior probability of inclusion (PrPI) decreased to 10\% ($a_\varphi = a_\zeta = 1$ and $b_\varphi = b_\zeta = 9$) and 1\% ($a_\varphi = a_\zeta = 1$ and $b_\varphi = b_\zeta = 99$).  We additionally observed lower specificity and higher sensitivity with increased prior probability of inclusion, but overall the differences were marginal. We found no evidence of sensitivity to $\sigma^2_{\beta_\theta}$ and   $\sigma^2_{\beta_\gamma}$ in terms of selection performance.

\section{Application}\label{application}
We apply our proposed method to analyze a microbiome  data set collected to study the relation between dietary intake and the human gut microbiome \citep{wu2011linking}. The data used in this analysis consist of 28 genera-level operational taxonomic unit counts obtained from 16S rRNA sequencing and a corresponding set of 97 dietary intake covariates derived from food frequency questionnaire on 98 subjects, resulting in over 2,500 potential relations between covariates and taxon abundances as well as zero-inflation indicators. Dietary covariates were standardized prior to analysis. In this data set, over 30\% of the observed reads were zeros, ranging from 0\% to roughly 75\% for each microbial taxon. 

In this analysis, we assumed a non-informative beta-binomial prior for inclusion indicators at both levels of the model ($a_\varphi = b_\varphi = a_\zeta = b_\zeta = 1$) and weakly-informative priors for regression coefficients ($\sigma^2_{\beta_\theta} = \sigma^2_{\beta_\gamma} = 5$). The MCMC algorithm was run for 10,000 iterations. After a burn-in of 5,000 samples, inference was drawn from the remaining 5,000 iterations, thinning to every $10^{th}$ iteration. Visual inspection of the trace plots for the number of active covariates in the model indicated good convergence and mixing. A covariate’s inclusion in the model was determined using the median model approach (i.e., MMPI $\geq 0.50$). Additionally, we compared the results to the variable selection methods discussed in Section \ref{simulations}.  
 
\begin{figure}[ht]
  \hspace*{-0.4cm}     
   \centering
       \includegraphics[scale=0.45]{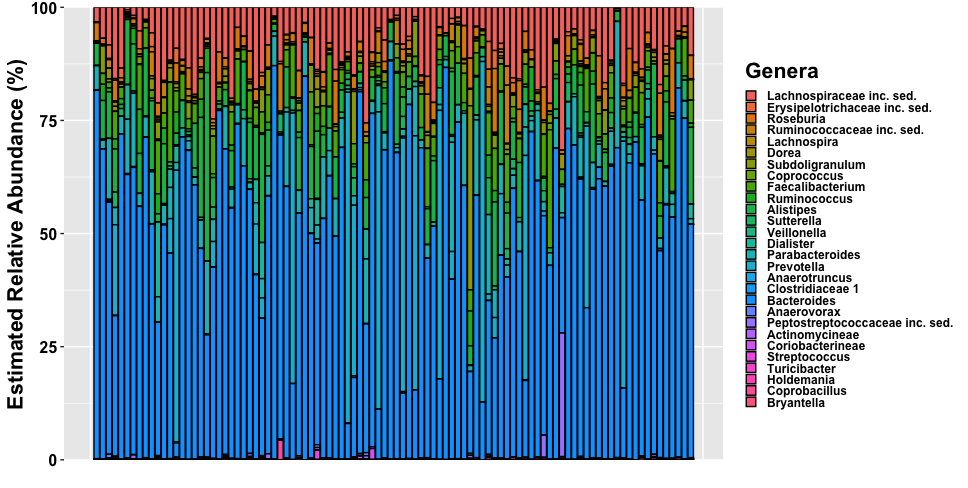} 
\caption{Application Results: Genus-level estimates of individual-level relative abundances for the application data with the proposed ZIDMbvs model.  \textit{inc.\ sed.}- incertae sedis.   }     \label{fig::abundances}
\end{figure}

 The individual-level relative abundances estimated by ZIDMbvs are presented in Fig. \ref{fig::abundances}. Since the true abundances are never known in practice, we compared the ZIDMbvs estimates to those obtained with the alternative methods assessed in Scenario 1 of the simulation study. We found that the models provided similar estimates overall, with ZIDMbvs, DMbvs, Tuyl's approach, and DirFactor the most similar (i.e., average absolute difference around 1$\times 10^{-4}$ and Frobenius norm around 0.02).

 \begin{figure}[ht]
 \hspace*{-1.3cm}
 \centering
       \includegraphics[scale=0.6]{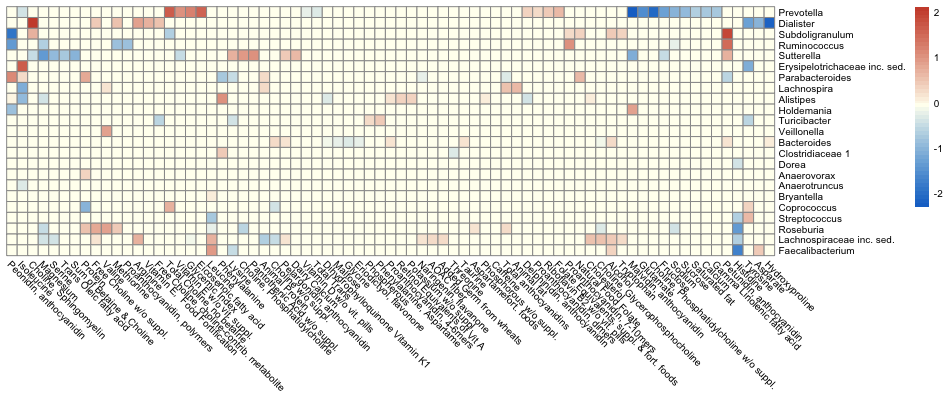}
 
    \caption{Application Results:  Dietary covariates identified as associated with relative taxa abundances using the ZIDMbvs model. \textit{inc.\ sed.}- incertae sedis.  }\label{fig::seleZIDM}
\end{figure}

 \textcolor{black}{Figure \ref{fig::seleZIDM} and Web Figures  S2, S3, and S4 present the dietary covariates identified as associated with relative taxa abundances using the ZIDMbvs, DMbvs, ZIDM, and DMpen  methods, respectively.}
With ZIDMbvs, we identified associations between dietary covariates and 23 of the 28 taxa. The highest concentration of associations were found with genera Prevotella (20), Lachnospiraceae Incertae Sedis (17), Bacteroides (14), and Sutterella (14). In previous studies, Prevotella and Bacteroides have been shown to be associated with high carbohydrate and protein/fat/choline diets, respectively. Similar patterns were observed with ZIDMbvs and DMbvs. In contrast to DMbvs, ZIDMbvs identified numerous associations between dietary intake and genus Sutterella, which has been linked to gastrointestinal diseases \citep{kaakoush2020sutterella}. Further, our proposed method identified 15 associations between dietary factors and the probability of an at-risk observation (Web Figure S5). Here, we found positive relations between carbohydrates (i.e., \textit{maltose} and \textit{added germ from wheats}) and an at-risk observation and a negative association between \textit{palmitelaidic trans fatty acid} and genus Prevotella. \textcolor{black}{  Compared to the proposed method, ZINB identified a similar number of associations, and the highest concentration of associations were with Prevotella. On the other hand, DMpen suggested a much sparser model and identified no associations with Prevotella. DMpen identified numerous relations with Bacteroides including positive associations with \textit{animal  and  dairy protein} as well as negative associations with \textit{maltose}, \textit{sucrose}, and \textit{added germ from wheats}. }
 
\textcolor{black}{To evaluate the methods' variable selection stability, we applied each method to 100 bootstrapped data sets generated from the application data and calculated the stability estimate, $\hat{\Phi}$, proposed by \cite{nogueira2017stability}, which ranges (asymptotically) from 0 to 1 with 1 indicating identical selection patterns across the bootstrap samples. \textcolor{black}{The proposed method obtained a relatively similar stability estimate ($\hat{\Phi}_{ZIDM} = 0.2399$) compared to the DM-based models ($\hat{\Phi}_{DMpen}$ $=0.1998$, p-value $= 0.11$; and  $\hat{\Phi}_{DM} = 0.1982$, p-value $ = 0.09$) and a higher stability estimate compared to ZINB ($\hat{\Phi}_{ZINB}$ $=0.0562$, p-value $< 0.001$). The corresponding p-values were obtained from a two-sided test comparing the variable selection stability of the proposed method and each of the competing methods following \cite{nogueira2017stability}.} Notably all of the methods obtained fairly poor stability estimates (i.e., $\hat{\Phi} \leq 0.40$ \citep{nogueira2017stability}), which may potentially reflect the large between-subject variability typically observed in human microbiome research studies. \textcolor{black}{ Additionally in the Supporting Information, we provide plots of the proportion of bootstrapped samples in which the associations identified in the application study were selected using each method (Figures S6-S9).} We provide a sensitivity analysis for the proposed method on the application data in the Supporting Information. }

\vspace{-1cm}
\section{Discussion}

In this work, we propose a  zero-inflated Dirichlet-multinomial model for multivariate compositional count data with excess zeros that provides both individual- and population-level inference without making restrictive assumptions or relying on approximation techniques. We then extend our model to regression settings and embed sparsity-inducing priors to perform variable selection for high-dimensional covariate spaces. In simulation, we demonstrate that our model is able to obtain similar estimation performance for population-level zero-inflation probabilities, population-level count probabilities, and individual-level count probabilities compared to existing methods. Notably, our approach is the only method to provide estimates for all of these measures, in addition to simultaneously estimating model uncertainty. \textcolor{black}{ While individual-level inference helps capture within- and between-subject heterogeneity critical for designing and evaluating personalized intervention strategies, population-level inference may help characterize the core microbiome, or a common set of taxa in a given host species or environment, which is a major goal in microbiome research studies \citep{turnbaugh2007human}. Additionally, the proposed method  is applicable to other settings that encounter zero-inflated multivariate compositional data in which population- and individual-level estimates may be of interest (e.g., biomedical and public health research, econometrics, and ecology). } We show that our method obtains better selection performance  than competing methods in various regression settings. Using a combination of multiple data augmentation techniques, our method is designed to scale to large compositional as well as covariate spaces while preserving inference. Further, our approach does not require burdensome tuning procedures for implementation and results were relatively robust to hyperparameter specification. We provide an \texttt{R}-package with a user-friendly vignette that implements the proposed ZIDM and ZIDMbvs models, in addition to the DM model, DMbvs model, and Tuyl's approach for parameter estimation and uncertainty quantification. The vignette contains a step-by-step tutorial demonstrating how to apply the proposed methods on simulated data as well as the gut microbiome data set analyzed in the application study.  

The development of a zero-inflated Dirichlet-multinomial model creates ample opportunity for future extensions that will enable more robust analysis of multivariate compositional count data found within and beyond omics research. It is well known that one of the major limitations of the DM distribution is that it does not account for positive and negative correlation structures among counts. However, the DM distribution is essential to the construction of a Dirichlet-tree multinomial distribution, which is able to accommodate more complex correlation structures. \textcolor{black}{Alternatively, the model could be developed with latent factors, similar to \citep{ren2017bayesian}, to accommodate more complex correlation structures among compositional abundances and potentially improve estimation performance.} While we showcased the proposed ZIDM distribution's flexibility to handle high-dimensional regression settings, future work could explore the use of the ZIDM distribution as a prior distribution to learn underlying latent structure in hierarchical models.

    \bibliographystyle{biom} 
 \bibliography{bib.bib}

\end{document}